\RequirePackage{fix-cm}
\documentclass{article}       
\usepackage{graphicx}
\usepackage{amsmath}%
\usepackage{amsfonts}%
\usepackage{amssymb}%
\usepackage{graphicx}
\usepackage[open,openlevel=2]{bookmark}

\newcommand{\Schrodinger}{Schr\"{o}dinger}
\begin{document}

\title{Relativity and Equivalence in Hilbert Space: \\A Principle-Theory Approach to the Aharonov-Bohm Effect\thanks{Forthcoming in \emph{Foundations of Physics}: \url{https://doi.org/10.1007/s10701-020-00322-y}}}

\author{Guy Hetzroni\thanks{guy.hetzroni@mail.huji.ac.il}\\ \normalsize Edelstein Center For the History and Philosophy of Science \\ \normalsize The Hebrew University of Jerusalem \\ \normalsize and \\\normalsize Department of Natural Sciences, The Open University of Israel.}

\date{}

\maketitle

\begin{abstract}
This paper formulates generalized versions of the general principle of relativity and of the principle of equivalence that can be applied to general abstract spaces. It is shown that when the principles are applied to the Hilbert space of a quantum particle, its law of coupling to electromagnetic fields is obtained. It is suggested to understand the Aharonov-Bohm effect in light of these principles, and the implications for some related foundational controversies are discussed. 
\end{abstract}

\section{Introduction}
\label{intro}
The clarity and elegance of the description of the gravitational interaction in the general theory of relativity often seem to stand in contrast with the confusion which surrounds the description of interactions at the microscopic level through quantum theory and its successors. The workings of the requirement for general covariance (Einstein's general principle of relativity) and the principle of equivalence leave comparatively little room for vagueness concerning the role of the abstract mathematical structures in general relativity. In this paper I propose a generalized version of these two principles in terms of symmetry and non-symmetry transformations. When the principles are applied to the Hilbert space of a quantum particle, its law of coupling to electromagnetic fields is naturally obtained. 

As a central application I provide an analysis of the Aharonov-Bohm effect \cite{aharonovbohm1959}, an effect which remains controversial as it celebrates its 60th anniversary. I then focus on the role of topology in explaining the effect. In contrast to the central existing accounts of the effect which explain it in terms of potentials, fields or holonomies, the suggested account manifests the principle-theory approach described a century ago by Einstein \cite{einstein1919times}. 

\section{The principles}
\label{principles}

\subsection{Symmetries: A Generalized Principle of Relativity}
\label{relativity}
In the course of the second half of the eighteenth century coordinate systems became a major mathematical tool used by physicists to represent physical systems. Coordinate systems allowed a common ground on which different kinds of motions in different physical contexts can be described \cite{meli1993reference}. The foundational significance of reference frames became apparent when inertial frames were introduced by Ludwig Lange\cite{lange2014inertia} in 1885. Inertial frames replaced the heavy ontological commitment of Newtonian absolute space with a minimalistic, operationally defined, mathematical structure required in order to consistently apply Newton's laws\cite{SEPinertialframes}. The special theory of relativity further emphasized the foundational importance of inertial frames. The privileged set of inertial frames of reference now took the role of both obsolete notions of Newtonian absolute space and the luminiferous aether. 

Surprisingly, the next major breakthrough, the general theory of relativity, was based on the banishment of inertial frames from their privileged status. Einstein had described the motivation for this move \cite{einstein1919times}:
\begin{quotation}
What has nature to do with our coordinate systems and their state of motion? If it is necessary for the purpose of describing nature, to make use of a coordinate system arbitrarily introduced by us, then the choice of its state of motion ought to be subject to no restriction; the laws ought to be entirely independent of this choice (general principle of relativity).
\end{quotation}

At first glance this principle seems like anything but a solid foundation for a physical theory. It is motivated by philosophical \emph{a priori} reasoning, not by empirical evidence. On the contrary, it is even in conflict with the successful physical theories that predated general relativity, and also with later successful theories such as quantum field theories. 

The principle regards the role of coordinate systems as a matter of mere labeling that should make no difference. According to Earman \cite{earman1989ST} the principle can be regarded as a manifestation of a general symmetry principle: any space-time symmetry of a physical theory should also be a dynamical symmetry of the theory. If a theory assumes that one space-time point is not inherently different from another, then the laws of motion should not discriminate between them and not depend upon their labels. 

The generalization of the principle to spaces that are more abstract than space-time is straightforward. \newline
\emph{\textbf{The generalized principle of relativity}: The form of the dynamical law should not change under change of representation}. 

The principle is therefore relevant when physical states have different possible mathematical representations. The methodology it suggests is to attempt to change the theory in a way that extends its invariance properties under different transformations. 

\subsection{Non-Symmetries: A Generalized Principle of Equivalence}
\label{equivalence}

The mathematical notion of general covariance that follows from the general principle of relativity does not provide by itself a sufficient basis for a gravitational theory. Likewise, in the different context of gauge theories, several philosophers have noted that the mathematical requirement of local gauge covariance cannot be held a satisfactory justification for the derivation of interactions \cite{brown1998objectivity,teller2000gauge,redhead2003gauge,BenMenahem2012symmetry}. The analogy between these two cases was noted by Lyre \cite{lyre2000equivalence,lyre2001gauging,lyre2003phase}. He suggested to supplement the requirement for local gauge covariance with a generalized equivalence principle, by analogy with the way the principle of equivalence supplements the requirement for general covariance in general relativity. Lyre's generalized equivalence principle is formulated in analogy to the idea of equality of inertial and gravitational mass. Its role is to bridge the gap between mathematical gauge transformations and physical interaction fields. The generalized principle of equivalence I present here serves to fulfill the same role, but it is based on a different formulation of the principle of equivalence in GR, emphasizing its methodological role. It is an improved version of a methodological principle I recently presented in a previous paper\cite{hetzroni2019ghosts}. 

The principle of equivalence becomes useful exactly where the general principle of relativity is violated, i.e. for those transformations under which the dynamical law changes its form. Einstein's principle of equivalence sets off from the simplest case of non-symmetry transformations in special relativity, which is the transition to a uniformly accelerating frame of reference. 

\begin{quotation}
``Can the principle of relativity be extended also to reference systems, which are (uniformly) accelerated relative to one another? The answer runs: As far as we really know the laws of nature, nothing stops us from considering the [accelerating] system $K’$ as at rest. If we assume the presence of a gravitational field (homogeneous in the first approximation) relative to $K’$.[...] The assumption that one may treat $K’$ as at rest with all strictness without any laws of nature not being fulfilled with respect to $K’$ I call the principle of equivalence.''\footnote{From Einstein's reply to Friedrich Kottler, Annalen der Physik 51:639-642 (1916) as, quoted in \cite{norton1985equivalence}. \label{kottler}}  
\end{quotation}

This simple ``elevator'' version of Einstein's principle of equivalence can also be presented from a slightly different point of view. Consider a localized mechanical system. In special relativity the processes in the system would be described from the perspective of some preferred (inertial) reference frame by certain equations. The same processes would be described from the perspective of other (non-inertial) frames of reference by modified equations; the general principle of relativity is thus violated. The principle of equivalence provides a way to get rid of this violation by modifying the original theory. For this purpose the principle conjectures the existence of an interaction which modifies the original equations that describe this localized system \emph{in exactly the same way} as does the non-inertial change of representation. Empirical considerations identify this conjectured interaction with the effect caused by a uniform gravitational field. 

The two principles therefore work together. In order to satisfy the general principle of relativity and have a theory that is invariant under \emph{passive} change of coordinates, the theory must include a parallel notion of \emph{active} change\footnote{To avoid ambiguities, it is emphasized that the terms `passive' and `active' are used throughout this paper to distinguish between transformations of the mathematical representation (that do not alter the physical state), and transformations which replace one physical configuration with a different one.\label{passiveactive}}. In the case of general relativity this is an active change with respect to the metric field, which is the physical object that is described by coordinate systems. The gravitational interaction instantiates this active transformation. 

From this perspective the principle of equivalence appears as a powerful epistemological principle: the mathematical scheme for moving from one representation to another in the interaction-free theory informs us of the form of the law of interaction. We therefore suggest the following methodological principle. \newline
\emph{\textbf{Generalized principle of equivalence}: For every possible change of the form of the dynamical law under change of representation, there is an interaction that is locally manifested as a change of an equivalent form.} 

It is important to note that while every change of representation is conjectured to have a parallel active change, the opposite is not necessarily true. The early controversies regarding the principle of equivalence in general relativity were in large part due to the fact that the effect of gravity in the theory is generally \emph{not} equivalent to the effect of a change of frame of reference. Spacetime curvature cannot be removed by a transformation of the coordinates. Thus, despite the notion of equivalence between coordinate transformations and some gravitational effects, the introduction of the gravitational interaction enriches the theory with new phenomena that cannot be regarded as a different description of interaction-free situations. Einstein emphasized in this context that ``one may in no way assert that gravitational fields should be explained so to speak purely kinematically [...] Merely by means of acceleration transformations from a Galilean system into another, we do not become acquainted with \emph{arbitrary} gravitational fields, but those of quite a special kind. [...] This is only again another formulation of the principle of equivalence.''\footnote{See footnote \ref{kottler}.}  We shall see that the same is true for the generalized principle. In general, neither the kinematics nor the dynamics of the interaction would be uniquely determined by the generalized principle of equivalence. 

The generalized principle of equivalence therefore establishes a methodology that adds an interaction to an interaction-free theory. The form of the law of interaction is derived based on a generalization of the active counterparts of the non-symmetries of the interaction-free theory.

\section{Applying the principles in Hilbert space and the AB effect}
\label{hilbert}
Various reconstructions of quantum theory set off from principles about information or locality, from which they derive some properties of the quantum structure. The approach presented here is different, and is arguably closer to the principle-theory approach presented by Einstein. I do not attempt to reconstruct quantum theory from scratch. Instead, the two generalized principles are applied in the given Hilbert space of a free quantum particle to get the interaction. 

\subsection{A quantum particle on a ring}
\label{ring}
We begin with a simple case. Consider a quantum particle of mass $m$ that is confined to a thin ring of radius $R$. The position basis can be parametrized by the polar angle $\theta$, such that the position eigenstates are written as $\vert\theta\rangle$. The momentary state $\vert\Psi\left(t\right)\rangle$ of the particle can be expressed in terms of the wavefunction $\Psi\left(\theta,t\right)=\langle\theta\vert\Psi\left(t\right)\rangle$. The angular momentum operator can be written in terms of a derivative of the wave function: $\hat{L}=-i\hbar \partial/\partial\theta$. The Hamiltonian is $\hat{H}=\hat{L}^2/2mR^2$, and the \Schrodinger\ equation is
\begin{equation}
\label{ringschrodinger}
i\hbar\frac{\partial\Psi}{\partial t}=-\frac{\hbar^2}{2mR^2}\frac{\partial^2\Psi}{\partial\theta^2}. 
\end{equation}
The eigenfunctions are $\psi_n\left(\theta\right)=\frac{1}{\sqrt{2\pi}}e^{i n\theta}$, and the energy spectrum is given by:
\begin{equation}
\label{ringenergies}
\epsilon_n=\frac{\hbar^2 }{2mR^2}n^2.
\end{equation}

On the backdrop of these familiar consideration it is now possible to apply the two principles. In order to do this we first need to ask what are the possible different representations of the states of the particle, and what would count as a change of representation. In quantum mechanics the answer is obvious: the different representations for the states of the systems are different bases of the Hilbert space. The transformations between the different representations are passive unitary transformations. They replace a representation in one basis with a representation in another basis. For every unitary transformation there is a third basis which is a preferred basis for the transformation in the sense that in this basis the transformation takes a diagonal form. That means that the variables changed in the transformation are the relative phases between the basis states. 

In this example we limit ourselves to those unitary transformations which are diagonal in position basis. These transformations take the form
\begin{equation}
\label{ringunitary}
\vert\theta\rangle\rightarrow \vert\theta'\rangle=U\vert\theta\rangle=e^{-i\Lambda\left(\theta\right)}\vert\theta\rangle. 
\end{equation} 
The wave function transforms accordingly: $\Psi\left(\theta,t\right)=\langle\theta\vert\Psi\rangle\rightarrow\Psi'\left(\theta,t\right)=\langle\theta'\vert\Psi\rangle=e^{i\Lambda\left(\theta\right)}\Psi\left(\theta,t\right)$. The transformation law of the operators is based on the fact that every linear operator $\hat{B}$ can be expressed as a linear combination of matrix elements of the form $\vert\theta_1\rangle\langle\theta_2\vert$, and therefore transforms as $B\rightarrow B'=UBU^\dagger$. The angular momentum operator therefore transforms as 
\begin{equation}
\label{ringLtrans}
\hat{L}\rightarrow\hat{L}'=U\hat{L}U^\dagger=-i\hbar\frac{\partial}{\partial\theta}-\omega\left(\theta\right), 
\end{equation}
with: 
\begin{equation}
\label{ringomega}
\omega\left(\theta\right)\equiv-\hbar\frac{\partial\Lambda}{\partial\theta}. 
\end{equation}

Thus, the dynamics is given by the \Schrodinger\ equation \eqref{ringschrodinger} only for a preferred representation in which $\hat{L}=-i\hbar\frac{\partial}{\partial\theta}$. Paul Dirac had similarly noted the many possible representations of the momentum operators \cite{dirac1958principles}. To the preferred representation (in which $p_j=-i\hbar\frac{\partial}{\partial x_j}$) he had referred as ``\Schrodinger's representation''. 

It is straightforward to see that this situation does not satisfy the generalized principle of relativity. The dynamical law takes a preferred form in a given representation, with an additional term added in other representations. (This form is analogous to the way classical dynamics takes a preferred form in inertial frames.) 

In order to extend the theory into a theory that is invariant under \eqref{ringunitary}, we apply the generalized principle of equivalence. The effect of a change of frame of reference is given by \eqref{ringLtrans}. We therefore postulate the existence of an interaction that locally takes a similar form:
\begin{equation}
\label{ringLactivetrans}
\hat{L}\rightarrow -i\hbar\frac{\partial}{\partial\theta}-\tilde{\omega}\left(\theta\right). 
\end{equation}

The transformed \Schrodinger\ equation is therefore:
\begin{equation}
\label{ringschrodingeromega}
i\hbar\frac{\partial\Psi}{\partial t}=\frac{1}{2mR^2}\left(-i\hbar\frac{\partial}{\partial\theta}-\tilde{\omega}\left(\theta\right)\right)^2\Psi. 
\end{equation}

The term equivalence refers to the similar mathematical form of the two transformations \eqref{ringLtrans} and \eqref{ringLactivetrans}, but it is clear that despite the similarity they represent  different things. The transformation \eqref{ringLtrans} is a passive transformation which represents the same angular momentum operator in the transformed basis. The transformation \eqref{ringLactivetrans} replaces the angular momentum operator with a local relation between the angular momentum and a physical (classical) field. This relation now takes the role of a dynamic variable in the Hamiltonian. The temporal evolution is therefore an active change of the state of the particle with respect to the field. 

The difference between the two transformations is not merely interpretational; they are also formally different. The transformation \eqref{ringLtrans} was obtained out of a well defined change of basis \eqref{ringunitary} parametrized by the single valued function $\Lambda\left(\theta\right)$. From this it follows that $\oint_{0}^{2\pi}\omega\left(\theta\right) d\theta=0$. The natural generalization would be to remove this constraint, and regard the physical field $\tilde{\omega}$ as arbitrary. The simplest case of physical importance is that in which it is constant. In this case we can write: $\tilde{\omega}=\frac{q\Phi}{2\pi}$ with $q$ a coupling constant, and the \Schrodinger\ equation \eqref{ringschrodingeromega} yields the energy spectrum: $\epsilon_n=\frac{\hbar^2 }{2mR^2}\left(n-\frac{q\Phi}{2\pi\hbar}\right)^2$, which is empirically distinguishable from \eqref{ringenergies} \cite{viefers2004quantum}. 

These equations correctly describe the influence of a classical magnetic field on the ring, with $\Phi$ the magnetic flux  through the ring and $q$ the charge of the particle. The shift of the energy levels is the static version of the Aharonov-Bohm effect, in which the effect depends on the total flux \emph{through} the ring, and is observable even if the magnetic field vanishes \emph{on} the ring.

\subsection{A quantum particle coupled to electromagnetic field}
\label{singleparticle}
In this subsection we start from the description of a single free quantum particle, and obtain the law of coupling of the particle to electromagnetic influence using the generalized principles. The derivation is similar to the case presented in the previous subsection. 

The changing state of a quantum particle is represented by the time dependent spatial wave function $\Psi\left(\vec{r},t\right)=\langle\vec{r}\vert\psi\left(t\right)\rangle$. A time dependent unitary transformation that is diagonal in position basis is given by a phase transformation that depends on both space and time:
\begin{equation}
\label{unitarytrans}
\vert\vec{r}\rangle\rightarrow \vert\vec{r}'\rangle=U\vert\vec{r}\rangle=e^{-i\Lambda\left(\vec{r},t\right)}\vert\vec{r}\rangle. 
\end{equation} 
Momentum operators transform: 
\begin{equation}
\label{Ptrans}
\hat{p_i}\rightarrow\hat{p_i}'=U\hat{p_i}U^\dagger=p_i-\omega_i\left(\theta\right)=-i\hbar\frac{\partial}{\partial x_i}-\omega_i\left(\theta\right), 
\end{equation}
with $\omega_i\equiv -\hbar\frac{\partial}{\partial x_i}\Lambda\left(\vec{r},t\right)$. And the energy operator: 
\begin{equation}
\label{Etrans}
\hat{E}\rightarrow\hat{E}'=U\hat{E}U^\dagger=i\hbar\frac{\partial}{\partial t}+\omega_0\left(\vec{r},t\right), 
\end{equation}
with $\omega_0\equiv -\hbar\frac{\partial}{\partial t}\Lambda\left(\vec{r},t\right)$. 

The \Schrodinger\ equation of a free particle 
\begin{equation}
\label{freeschrodinger}
i\hbar\frac{\partial}{\partial t}\Psi\left(\vec{r},t\right)=-\frac{\hbar^2}{2m}\nabla^2\Psi\left(\vec{r},t\right) 
\end{equation}
transforms to the equation 
\begin{equation}
\label{schrodingeromega}
\left(i\hbar\frac{\partial}{\partial t}+\omega_0\left(\vec{r},t\right)\right)\Psi\left(\vec{r},t\right)=\frac{1}{2m}\left(-i\hbar\nabla-\vec{\omega}\left(\vec{r},t\right)\right)^2\Psi\left(\vec{r},t\right)
\end{equation}
(with $\vec{\omega}$ for the vector field whose Cartesian components are $\omega_i$). 

The theory does not satisfy the generalized principle of relativity due to this non-invariance. Like in the previous subsection, we apply the generalized principle of equivalence by introducing an interaction with an external field in order to obtain an extended invariant theory. The effect of the field on the particle is locally equivalent to a change of representation, therefore the energy and momentum operators transform as
\begin{align}
\label{PEtransactive}
p_i &\rightarrow p_i - \tilde{\omega}_i    &    E &\rightarrow E + \tilde{\omega}_0.
\end{align}
The transformed \Schrodinger\ equation is therefore 
\begin{equation}
\label{schrodingeromegaactive}
\left(i\hbar\frac{\partial}{\partial t}+\tilde{\omega}_0\left(\vec{r},t\right)\right)\Psi\left(\vec{r},t\right)=\frac{1}{2m}\left(-i\hbar\nabla-\vec{\tilde{\omega}}\left(\vec{r},t\right)\right)^2\Psi\left(\vec{r},t\right).	
\end{equation}

The corresponding physical fields are identified when empirical considerations are taken into account: Equation \eqref{schrodingeromegaactive} is identical to the \Schrodinger\ equation coupled to electromagnetism that is obtained using standard minimal coupling procedure if $\tilde{\omega}=\frac{q}{c}\vec{A}\left(r,t\right)$ and $\tilde{\omega}_0=-qV\left(r,t\right)$. 

Equation \eqref{schrodingeromegaactive} is invariant under the transformation
\begin{align}
\label{gaugetrans}
\vert\vec{r}\rangle&\rightarrow e^{-i\Lambda\left(\vec{r},t\right)}\vert\vec{r}\rangle & \vec{A}&\rightarrow\vec{A}-\frac{\hbar c}{q}\nabla\Lambda\left(\vec{r},t\right) & V\left(\vec{r},t\right)&\rightarrow V\left(\vec{r},t\right)+\frac{\hbar}{q}\frac{\partial}{\partial t}\Lambda\left(\vec{r},t\right). 
\end{align} 
The theory now satisfies the generalized principle of relativity if this transformation is regarded as one change of representation of the state of both the particle and the field. The physically significant quantities on which the interaction depends are the relations $\vec{p}-\frac{q}{c}\vec{A}$ and $E-qV$. The above transformation is a change of representation in the sense that it does not change these quantities. 

Equation \eqref{schrodingeromegaactive} describes the change of a wave function that is coupled to external physical fields. It corresponds to the well known Hamiltonian which takes the two fields into account:
\begin{equation}
\label{EMschrodingerH}
H=\frac{1}{2m}\left(p-\frac{q\vec{A}}{c}\right)^2 + qV. 
\end{equation}

\subsection{The dynamical Aharonov-Bohm effect}
\label{ABeffect}

The principle of equivalence states that every change of representation has a parallel interaction that induces change of the same form, but not vice versa. A change in the curvature has no equivalent change of coordinates. This is also the case with the equivalence between local phase transformations and coupling to electromagnetic fields. One way to see it is to note that even though a change of basis does change the form of the momentum operators according to \eqref{Ptrans}, it does not change the commutation relation between them ($\left[p'_i,p'_j\right]=\left[p_i,p_j\right]$ as a direct result of the definition $\omega=-\hbar\vec{\nabla}\Lambda$). In contrast, the replacement in \eqref{PEtransactive} of $p_i$ with $\pi_i\equiv p_i-\frac{q}{c}A_i$ may generally change the commutation relations, since $A$ is arbitrary. It is easy to see that the relation $\left[\pi_i,\pi_j\right]$ is proportional to a component of the magnetic field. 

What about the case $\left[\pi_i,\pi_j\right]=\left[p_i,p_j\right]=0$? It may be tempting to think that if this condition holds in a given region of space, then there is no real electromagnetic influence on that region, as a simple change of basis would suffice to undo the transition from $p_i$ to $\pi_i$. The Aharonov-Bohm effect demonstrates that this is not true. The experiment demonstrates that an electron interference pattern can be altered by a static magnetic flux that only exists inside a shielded cylinder, where the wave function is zero. 

At this point many explanations of the effect (e.g. in \cite{ryder1996qft}, and see also Section \ref{conclusion}) turn to emphasize what appears to be the essential part of topological considerations. The effect, according to this approach, only exists because of the non-trivial topology of the configuration space of the particle (understood as the region of space in which the wave function can get non-zero values), and is explained by it. In this kind of explanation the situation is described from the point of view of an observer who has no access to the inside of the cylinder, where the flux is. As long as the observer limits herself to experiments that are conducted in contractible regions of the configuration space, she will not be able to tell whether there is a magnetic flux in the cylinder. The effect would only appear once two wave packets go around the cylinder, forming a domain that is not simply connected. 

According to the view that is presented here, the equivalence between the effect of an interaction and local phase transformation is formal; they are not the same thing. Accordingly, the role of topology is descriptive, not fundamental. The heart of the matter is the existence of the small region in which $p$ is replaced by $\pi$ such that $\left[\pi_i,\pi_j\right]\ne \left[p_i,p_j\right]$. This ensures that one physical situation is replaced by a different one. The new situation cannot be regarded as mere different description of the old situation. The principle of equivalence implies that \emph{locally}, in some other regions of space-time, the interaction would appear like a change of representation. The fiber-bundle formalism expresses this local equivalence in the possibility of mapping the solutions of the \Schrodinger\ equations in patches of spacetime to solutions of the interaction-free equation in these regions. This possibility is useful for the calculation of the phase shift. It should not be regarded, however, as the reason for it.  

It is true that any experiment that is confined to a simply-connected region of space in which $\left[\pi_i,\pi_j\right]= \left[p_i,p_j\right]$ cannot reveal the presence of a magnetic field outside that region. But this fact does not mean that the remote field does not change the physical situations in this region. The Aharonov-Bohm experiment, performed in a domain that can be regarded as a union of two such simply-connection regions, demonstrates that the relation between the physical situation in one simply-connected region and the physical situation in the other has definitely changed due to the electromagnetic interaction. 

\section{Conclusion}
\label{conclusion}

\subsection{Ontology}
\label{conclusionontology}

The Aharonov-Bohm effect serves to widen the spectrum of possible interpretations of the theory of electromagnetism beyond particles and fields\footnote{A comprehensive discussion of the different interpretations is given in \cite{belot1998EM}.}. One far-reaching possibility that was raised in the paper by Aharonov and Bohm \cite{aharonovbohm1959} is that of taking the electromagnetic potentials seriously as a fundamental physical entity and to ``define the physical difference between two quantum states which differ only by gauge transformation'' (p. 491). Later, Wu and Yang \cite{wu1975factors} proposed to understand electromagnetic interactions in terms of holonomies. Another approach suggests to regard the fiber-bundle structure as the fundamental physical entity \cite{Nounou2003AB,arntzenius2014space}. These approaches appeal to some hidden structure they postulate to exist in the world, resembling what Einstein had called a constructive theory approach\cite{einstein1919times}. To conclude, I would like to note that the suggestion to replace the constructive approach by a principle-theory approach does not mean total renunciation of the aim to explain the observed phenomena by appealing to physical properties of the world. Once the success of the generalized principles of relativity and of equivalence has been established, we are in a position to ask about the underlying ontological properties. What features of the physical world are reflected in the fact that physical situations have many possible mathematical representations? Why does every change of representation have a corresponding local change in the world that can be realized in an interaction? 

A possible answer to both questions can be found in the understanding of gauge theories as a manifestation of the relational nature of physical quantities \cite{rovelli2014gauge,hetzroni2019ghosts}. The generalized principle of relativity reflects the claim that the laws of physics should depend on the relations between fundamental physical objects, and not on the relations of the physical objects to a fictitious mathematical frame of reference. The generalized principle of equivalence is valid because for every possible change of a relation between a physical object and the mathematical frame of reference which is used to describe its state, there is a corresponding possible change of the relation between the physical object and another physical object. Interactions cause actual changes of the latter type. This correspondence is reflected in the structure of gauge theories when the formal role of a connection (a generalization of the concept of a frame of reference) is taken by an actual physical field that describes the physical configuration. 

The gauge invariant quantity that seems to represent a relation between the field and the particle is the canonical momentum $\pi^{\mu} = p^{\mu}-\frac{q}{c}A^{\mu}$. In contrast to the gauge invariant electromagnetic fields and to the holonomies, both characterize the electromagnetic situation, the canonical momentum is a genuine relational quantity between the two objects. The choice of a given gauge amounts to a choice of phase convention which divides this quantity into the phase gradient of the particle and the electromagnetic potential, but there seems to be no reason to regard this division as more than conventional. The observable properties of the particle depend on it, rather then on the momentum $p^\mu$ itself. On the electromagnetic side, the components of the electromagnetic tensor $F^{\mu\nu}$ can be derived from the canonical momentum through its commutation relations. 

The principle-theory approach to the Aharonov-Bohm effect thus reveals a new way to think about the ontology of the electromagnetic interaction. The matter field and the interaction field represent two distinct physical objects, but the interaction between the two depends on only one fundamental dynamical physical quantity $\pi^\mu$, which is the relation between them. 

\subsection{Fiber Bundles, Topology and the Applicability of Mathematics}
\label{conclusiontopology}

The structure of general relativity provided the inspiration for the first attempt, by Weyl \cite{weyl1918gravitation}, to formulate electromagnetism in terms of gauge. Soon after the introduction of \Schrodinger's theory, it was realized by London \cite{london1927gauge} that Weyl's ideas can be formulated in terms of quantum phases, which had lead to Weyl's second theory \cite{weyl1929elektron}. Weyl showed that the transition from global to local phase invariance can facilitate the law of electromagnetic coupling of the electron, in analogy with the transition from the special theory of relativity to the general theory. In developing his theories Weyl was motivated by a philosophical view that favors pure \emph{a priori} mathematical and geometrical considerations as the basis for theories, and considered local symmetry transformations as the ultimate expression of such considerations \cite{ryckman2003gauge}. Some surprising consequences of the electromagnetic gauge theory were only revealed later, in the form of the Aharonov-Bohm effect. The common topological approach to the effect can be easily seen as based on an aim that may be similar to Weyl's, to explain physical phenomena in mathematical terms. Thus, continuous symmetries and topological considerations raise in a related way the question of the applicability of mathematics to physics. 

Wigner \cite{wigner1990effectiveness} has seen the effectiveness of mathematics in physics as an incomprehensible miracle, referring in particular to the adequacy of abstract  structures such as the Hilbert space representation in quantum mechanics. Steiner \cite{steiner1998applicability} provided a detailed philosophical analysis of the question, in which gauge symmetries and the issue of topology are brought as central examples. In these cases (and in some others) Steiner is astonished by the success of equations and theories that are constructed based on formal analogies, and sees it as a support for anthropocentric, non-naturalist philosophical view. 

This paper, on the one hand, supports Weyl's analogy by pointing out that the electromagnetic interaction of the quantum particle manifests the same principles that are manifested by the general theory of relativity, applied to a different state-space. The generalized principle of relativity repeats a common view on the issue, which goes back to Weyl. The formulation presented here emphasizes that it is an \emph{a priori} principle, and that it applies to the theoretical representation of the world, not to the world itself.

But on the other hand, the generalized principle of equivalence proposed here goes beyond Weyl's theory, and suggests a very different relation between the physical and the mathematical than the one that was pursued by Weyl. The mathematical transformation which relabels the states is not in itself physically significant, it is the corresponding active transformation that takes an identical form and represents an actual change in the world. The existence of such a possible change, and the existence of an object with respect to which the change can take place, has to be tested empirically. There is no \emph{a priori} reason why would the gravitational mass equal the inertial mass, nor for the locally similar structure of interactions and coordinate transformations. The principle of equivalence therefore expresses a kind of a guesswork. It is constructed at first with the purpose of satisfying the general principle of relativity, and its consequences have to be ultimately justified on empirical grounds. 

In general relativity coordinate transformations are directly equivalent to some ways in which the gravitational interaction can change space-time, not to the change caused by a general gravitational field. The analogy with electromagnetism is clear. A change of basis \eqref{unitarytrans} is not equivalent to the introduction of a general electromagnetic field, but only to a special case of it. This is because of the condition $\omega^\mu = -\hbar\partial^\mu\Lambda$ that is implicit in the definition of a change of basis. The general form of the interaction is obtained by omitting this condition in the transformation \eqref{PEtransactive}. Interestingly, it is exactly the Aharonov-Bohm effect that demonstrates that even those interactions that satisfy the condition, and are thus locally equivalent to a change of representation, are actual interactions that can make an observable difference, and should therefore be understood as changing the physical state.

Thus, the active view of the transformation also plays a crucial role in understanding the applicability of topological considerations. The point here is not the manner in which topology helps to understand the properties of the interaction\footnote{This issue is the subject of the lion's of the contemporary philosophical discourse about the topological approach to the effect,  see for example \cite{Nounou2003AB,shech2018idealizations} and references therein.}, but rather that the nature of the physical interaction, as it is expressed by the generalized principle of equivalence, helps us to understand the applicability of topology. 

Allowing the phase-convention to change according to \eqref{unitarytrans}, the generators of passive translations in the configuration space of the particle are the 3 operators $-i\frac{\partial}{\partial x_j}+\frac{\partial\Lambda}{\partial x_j}$. That means that passive infinitesimal translation $\vec{\Delta r}$ transforms $\psi\left(\vec{r},t\right)$ into $\psi\left(\vec{r}+\vec{\Delta r},t\right) e^{i\vec{\Delta r}\cdot\nabla\Lambda(\vec{t},t)}$. The function $\Lambda$ defines the change in the phase convention, and the second term in the generator of translations reflects the change of phase due to the local change of convention. The term convention is justified since the choice of $\Lambda$ makes no observable difference. In other words, $\nabla\Lambda$ amounts to a flat connection. Active translation is understood here as a transport of the wave function in space with respect to other physical objects (e.g. the solenoid)\footnote{see footnote \ref{passiveactive}}. This is an actual change in the state, and therefore the phase can change in a more general way, such that $\psi\left(\vec{r},t\right)$ would be transformed into $\psi\left(\vec{r}+ \vec{\Delta r},t\right) e^{i\vec{\Delta r}\cdot\vec\omega(\vec{r},t)}$, with $\omega$ a general function. The common analysis of the experiment regards the change of the wave packets as a sequence of such active transformations\footnote{This is an approximation assuming that the wave packets are fast, such that they do not change their form during the experiment.}. Here $\omega$ plays the role of a general connection, which is then reinterpreted as a representation of the magnetic field.

The difference-maker in the experiment is obviously the region of space in which the existence of the magnetic field introduces curvature to the connection of the $U\left(1\right)$ principal bundle over configuration space. The existence of this region implies that the entire situation is not equivalent to a free particle, and the existence of observable consequences (in the form of a shift of the interference pattern or a change in the energy levels) should therefore come as no surprise. The principle asserts that locally the interaction has the same form as a change of basis, not that it is globally equivalent to it. Topology and the fiber-bundle formalism allow for a convenient mathematical description which expresses this difference between local and global characteristics. It should not be regarded as more than a convenient mathematical description of the physical properties of the interaction. 

Symmetry and topology are mathematical notions whose applicability to physics is often portrayed as mysterious or even miraculous. The principles proposed in this paper aim at removing that mystery by establishing a connection between passive mathematical transformations and active changes in the world, and noting that while they share a similar form, they thoroughly differ in their meaning.

\section*{Acknowledgments}
I would like to thank Yemima Ben-Menahem and Daniel Rohrlich for their guidance and advice. This research was supported by the Israel Science Foundation (grant no. 1190/13) and by The Open University of Israel's Research Fund (grant no. 41240). This is a pre-print version of an article accepted for publication in \emph{Foundations of Physics}. The final authenticated version would be available online at: \url{https://doi.org/10.1007/s10701-020-00322-y}


\begin{thebibliography}{}
\providecommand{\url}[1]{{#1}}
\providecommand{\urlprefix}{URL }
\expandafter\ifx\csname urlstyle\endcsname\relax
  \providecommand{\doi}[1]{DOI~\discretionary{}{}{}#1}\else
  \providecommand{\doi}{DOI~\discretionary{}{}{}\begingroup
  \urlstyle{rm}\Url}\fi

\bibitem{aharonovbohm1959}
Aharonov, Y., Bohm, D.: Significance of electromagnetic potentials in the
  quantum theory.
\newblock Physical Review \textbf{115}(3), 485 (1959)

\bibitem{arntzenius2014space}
Arntzenius, F.: Space, time, and stuff.
\newblock Oxford University Press (UK) (2014)

\bibitem{belot1998EM}
Belot, G.: Understanding electromagnetism.
\newblock The British Journal for the Philosophy of Science \textbf{49}(4),
  531--555 (1998)

\bibitem{BenMenahem2012symmetry}
Ben{-}Menahem, Y.: Symmetry and causation.
\newblock Iyyun \textbf{61}, 193--218 (2012)

\bibitem{brown1998objectivity}
Brown, H.R.: Aspects of objectivity in quantum mechanics.
\newblock In: J.~Butterfield, C.~Pagonis (eds.) From physics to philosophy.
  Cambridge University Press (1999)

\bibitem{dirac1958principles}
Dirac, P.A.M.: The principles of quantum mechanics, 4th edn.
\newblock Oxford university press, London (1958)

\bibitem{SEPinertialframes}
DiSalle, R.: Space and time: Inertial frames.
\newblock In: E.N. Zalta (ed.) The Stanford Encyclopedia of Philosophy, winter
  2016 edn. Metaphysics Research Lab, Stanford University (2016)

\bibitem{earman1989ST}
Earman, J.: World Enough and Space-Time: Absolute Versus Relational Theories of
  Space and Time.
\newblock The MIT Press, Cambridge, Massachusetts (1989)

\bibitem{einstein1919times}
Einstein, A.: What is the theory of relativity.
\newblock The London Times  (1919).
\newblock Reprinted in \cite{einstein1954ideas} pp.227-32

\bibitem{einstein1954ideas}
Einstein, A.: Ideas and opinions.
\newblock Crown Publisher, New York (1954)

\bibitem{hetzroni2019ghosts}
Hetzroni, G.: Gauge and ghosts.
\newblock The British Journal for the Philosophy of Science  (2019).
\newblock Forthcoming. DOI:10.1093/bjps/axz021

\bibitem{lange2014inertia}
Lange, L.: On the law of inertia.
\newblock The European Physical Journal H \textbf{39}(2), 251--262 (2014).
\newblock Translation of ``{\"U}ber das Beharrungsgesetz'' (1885), translated
  by Herbert Pfister

\bibitem{london1927gauge}
London, F.: Quantenmechanische deutung der theorie von {W}eyl.
\newblock Zeitschrift für Physik A Hadrons and nuclei \textbf{42}, 375 (1927).
\newblock English translation, `Quantum-mechanical interpretation of {W}eyl’s
  theory' in \cite{ORaifeartaigh1997gauge}

\bibitem{lyre2000equivalence}
Lyre, H.: A generalized equivalence principle.
\newblock International Journal of Modern Physics D \textbf{9}(06), 633--647
  (2000)

\bibitem{lyre2001gauging}
Lyre, H.: The principles of gauging.
\newblock Philosophy of Science \textbf{68}(S3), S371--S381 (2001)

\bibitem{lyre2003phase}
Lyre, H.: On the equivalence of phase and field charges.
\newblock arXiv preprint hep-th/0303259  (2003)

\bibitem{meli1993reference}
Meli, D.B.: The emergence of reference frames and the transformation of
  mechanics in the enlightenment.
\newblock Historical studies in the physical and biological sciences
  \textbf{23}(2), 301--335 (1993)

\bibitem{norton1985equivalence}
Norton, J.: What was einstein’s principle of equivalence?
\newblock Studies in History and Philosophy of Science \textbf{16}(3), 203--246
  (1985)

\bibitem{Nounou2003AB}
Nounou, A.M.: A fourth way to the {A}haronov-{B}ohm effect.
\newblock In: K.~Brading, E.~Castellani (eds.) Symmetries in physics:
  philosophical reflections. Cambridge University Press (2003)

\bibitem{ORaifeartaigh1997gauge}
O'Raifeartaigh, L.: The dawning of gauge theory.
\newblock Princeton University Press (1997)

\bibitem{redhead2003gauge}
Redhead, M.: The interpretation of gauge symmetry.
\newblock In: K.~Brading, E.~Castellani (eds.) Symmetries in physics:
  philosophical reflections. Cambridge University Press (2003)

\bibitem{rovelli2014gauge}
Rovelli, C.: Why gauge?
\newblock Foundations of Physics \textbf{44}(1), 91--104 (2014)

\bibitem{ryckman2003gauge}
Ryckman, T.A.: The philosophical roots of the gauge principle: Weyl and
  transcendental phenomenological idealism.
\newblock In: K.~Brading, E.~Castellani (eds.) Symmetries in physics:
  philosophical reflections. Cambridge University Press (2003)

\bibitem{ryder1996qft}
Ryder, L.H.: Quantum field theory.
\newblock Cambridge university press (1996)

\bibitem{shech2018idealizations}
Shech, E.: Idealizations, essential self-adjointness, and minimal model
  explanation in the aharonov--bohm effect.
\newblock Synthese \textbf{195}(11), 4839--4863 (2018)

\bibitem{steiner1998applicability}
Steiner, M.: The applicability of mathematics as a philosophical problem.
\newblock Harvard University Press (1998)

\bibitem{teller2000gauge}
Teller, P.: The gauge argument.
\newblock Philosophy of Science \textbf{67}, S466--S481 (2000)

\bibitem{viefers2004quantum}
Viefers, S., Koskinen, P., Deo, P.S., Manninen, M.: Quantum rings for
  beginners: energy spectra and persistent currents.
\newblock Physica E: Low-dimensional Systems and Nanostructures \textbf{21}(1),
  1--35 (2004)

\bibitem{weyl1918gravitation}
Weyl, H.: Gravitation and electricity.
\newblock Sitzungsber. K{\"o}nigl. Preuss. Akad. Wiss. \textbf{26}, 465--480
  (1918)

\bibitem{weyl1929elektron}
Weyl, H.: Elektron und gravitation. i.
\newblock Zeitschrift f{\"u}r Physik A Hadrons and Nuclei \textbf{56}(5),
  330--352 (1929)

\bibitem{wigner1990effectiveness}
Wigner, E.P.: The unreasonable effectiveness of mathematics in the natural
  sciences.
\newblock In: Mathematics and Science, pp. 291--306. World Scientific (1990)

\bibitem{wu1975factors}
Wu, T.T., Yang, C.N.: Concept of nonintegrable phase factors and global
  formulation of gauge fields.
\newblock Physical Review D \textbf{12}(12), 3845 (1975)

\end{thebibliography}
\end{document}